\documentclass[conference]{IEEEtran}
\IEEEoverridecommandlockouts
\usepackage{cite}
\usepackage{amsmath,amssymb,amsfonts}
\usepackage{amsthm}
\usepackage{algpseudocode}
\algrenewcommand\alglinenumber[1]{} 
\usepackage{algorithm}
\usepackage{graphicx}
\usepackage{textcomp}
\usepackage{xcolor}
\newtheorem{lemma}{Lemma}
\newtheorem{theorem}{Theorem}
\usepackage{caption} 
\usepackage{dblfloatfix}
\usepackage[a4paper, top=0.7in, left=0.75in, right=0.75in, bottom=1in]{geometry}
\usepackage{algpseudocode}
\algtext*{EndIf}
\algtext*{EndFor}
\def\BibTeX{{\rm B\kern-.05em{\sc i\kern-.025em b}\kern-.08em
    T\kern-.1667em\lower.7ex\hbox{E}\kern-.125emX}}

\begin{document}

\title{Computing the Saturation Throughput for Heterogeneous p-CSMA in a General Wireless Network\\
}

\author{\IEEEauthorblockN{Faezeh Dehghan Tarzjani}
\IEEEauthorblockA{\textit{Dept. of Electrical and Computer Engineering} \\
\textit{University of Southern California}\\
Los Angeles, CA \\
dehghant@usc.edu}
\and
\IEEEauthorblockN{Bhaskar Krishnamachari}
\IEEEauthorblockA{\textit{Dept. of Electrical and Computer Engineering} \\
\textit{University of Southern California}\\
Los Angeles, CA \\
bkrishna@usc.edu}
}

\maketitle

\begin{abstract}
A well-known expression for the saturation throughput of heterogeneous transmitting nodes in a wireless network using p-CSMA, derived from Renewal Theory, implicitly assumes that all transmitting nodes are in range of, and therefore conflicting with, each other. This expression, as well as simple modifications of it, do not correctly capture the saturation throughput values when an arbitrary topology is specified for the conflict graph between transmitting links. For example, we show numerically that calculations based on renewal theory can underestimate throughput by 48–62\% for large packet sizes when the conflict graph is represented by a star topology. This is problematic because real-world wireless networks such as wireless IoT mesh networks are often deployed over a large area, resulting in non-complete conflict graphs. To address this gap, we present a computational approach based on a novel Markov chain formulation that yields the exact saturation throughput for each node in the general network case for any given set of access probabilities, as well as a more compact expression for the special case where the packet length is twice the slot length. Using our approach, we show how the transmit probabilities could be optimized to maximize weighted utility functions of the saturation throughput values. This would allow a wireless system designer to set transmit probabilities to achieve desired throughput trade-offs in any given deployment. 
\end{abstract}


\section{Introduction}

Carrier Sense Multiple Access (CSMA) is a core medium access mechanism in wireless networks, used in standards such as IEEE 802.11 and IEEE 802.15.4. It enables efficient sharing of communication channels, making it fundamental for the design of wireless protocols. A simplified version of CSMA, known as p-persistent CSMA (or p-CSMA, for short), first proposed by Kleinrock and Tobagi~\cite{kleinrock1975packet}, involves users synchronously contending for access to the communication medium using a fixed probability $p$ whenever the medium is idle. Due to its memoryless nature, this version is more analytically tractable, enabling it to be widely used in theoretical studies. Previous research has shown that $p$-persistent CSMA closely approximates the CSMA protocol used in IEEE 802.11 when $p$ is selected to align with the average backoff intervals, making it a robust model for understanding the performance of real-world networks~\cite{bianchi2000performance, cali2000dynamic, bruno2002optimization}. 
The studies mentioned above all rely on a uniform model where all nodes use the same access probability.


The heterogeneous $p_i$-persistent CSMA algorithm, which allows users to adopt distinct access probabilities $p_i$, allows for more nuanced performance characterization and leads to a throughput region that captures all feasible throughput combinations. Motivated by standards such as IEEE 802.11E, this model has been explored to optimize channel access in diverse network settings~\cite{menache2011reservation, inaltekin2008analysis, hsu2010analysis}.

In a heterogeneous $p_i$-persistent CSMA network with
$n$ backlogged users, transmissions occur at the beginning of each time slot, and a transmission is successful only if no other user transmits simultaneously. The saturation throughput, a key performance metric for CSMA-based networks, quantifies the long-term average rate of successful transmissions per user, focusing on contention rather than traffic patterns. Renewal theory~\cite{ross2014introduction}, has been widely used to approximate saturation throughput under the assumption that all transmitting nodes interfere with each other~\cite{gai2011saturation}. This approach simplifies the analysis but limits its applicability to real-world distributed wireless networks such as IoT mesh networks where nodes are likely to be spatially distributed over a wide area~\cite{cilfone2019wireless}.

In practice, such real-world wireless mesh networks exhibit heterogeneous connectivity, influenced by transmission power, path loss, and spatial distribution. This complexity can be modeled using a general conflict graph, where the nodes represent the transmitters, and the edges denote the interference constraints~\cite{li2017conflict}. 
Conflict graphs capture localized contention, allowing nonadjacent nodes to transmit independently while neighboring nodes must adjust their backoff processes. This approach enables a more realistic analysis of CSMA-based wireless networks, considering spatial reuse, partial contention, and intricate dependencies between competing transmitters.

However, to our knowledge, no previous work has addressed the problem of exactly computing saturation throughput in a heterogeneous $p_i$-CSMA wireless network with an arbitrary conflict graph $G$. Using the classic approach based on renewal theory, even if adapted to an arbitrary conflict graph, results in poor approximations if the conflict graph topology is sparse. For example, through numerical simulations we show that a renewal-theory-based approach can underestimate throughput by 48–62\% for large packet sizes when the conflict graph is represented by a star topology. Using an inaccurate approach to estimating throughput can in turn result in suboptimal settings of transmit probabilities, significantly affecting the performance of a given wireless network.

This paper addresses the problem of computing the \emph{exact} saturation throughput in a general wireless network with an arbitrary conflict graph $G$, where each transmitter follows a heterogeneous $p_i$-CSMA protocol. By accurately computing exact throughput values, our analysis further exposes interference asymmetry, enabling precise optimization of sensing and backoff parameters. This improves network capacity by mitigating contention and improves energy efficiency through collision reduction and adaptive transmission control, making it essential for optimizing real-world wireless networks.
\\

\noindent \textbf{Key Contributions.} The main contributions of this paper can be summarized as follows:
\begin{enumerate}
  \item We present a novel computational approach based on a Markov chain model, providing exact Saturation throughput values for each node in arbitrary contention graphs, extending beyond the limitations of existing expressions that assume complete contention.
  \item For the special case where the packet length is twice the slot length, we derive a compact, computationally efficient solution that maintains exactness while reducing complexity.
  \item We validate our approach by demonstrating its accuracy through comparison with Monte Carlo simulations, highlighting its accuracy compared to previous incorrect approximations.
  \item We illustrate how to optimize weighted sums of utility functions of saturation throughput using our solution, enabling efficient exploration of the trade-offs in network performance. 
  \item By being able to accurately predict throughput as a function of each node’s transmit probability, our framework enables node-by-node optimization to significantly increase overall network throughput or improve throughput fairness across heterogeneous wireless links. 
\end{enumerate}

These contributions collectively provide a rigorous framework for analyzing and optimizing throughput metrics in heterogeneous CSMA wireless networks in the general case, which will potentially leading to a positive impact on practical deployments.

The remainder of the paper is organized as follows: Section 2 provides an overview of the existing literature in related fields. Section 3 establishes the theoretical foundations necessary for our network model using conflict graphs and defines the saturation throughput metric under heterogeneous p-persistent CSMA. Section 4 discusses Renewal Theory and its shortcomings. Section 5 elaborates our exact Markov chain formulation and algorithm for throughput computation for a general T; for T=2, we introduce closed-form steady-state probabilities using detailed balance. Additionally, Section 5 develops an optimization framework for weighted throughput maximization using gradient ascent on transmission probabilities. Section 6 presents empirical validation, comparing our results with Monte Carlo simulations and classical models. Finally, Section 7 concludes the paper by summarizing our findings and suggesting directions for future research.

\section{Related Works}

Leith et al.~\cite{leith2009log, leith2010utility} study how to achieve utility-based fairness in wireless mesh networks using 802.11 devices. They show that the rate region is log-convex, extending previous results for Aloha networks, and use this property to find utility-fair solutions for a specific family of utility functions. In addition, they characterize the max-min fair solution and prove the convexity of the non-achievable rate region, enabling utility fairness for a broader set of utility functions. The authors of \cite{subramanian2013rate} characterize the rate region for CSMA/CA WLANs. This characterization not only sheds light on the fundamental properties of CSMA/CA WLANs, but also enables the application of convex optimization techniques to address utility-fair resource allocation problems in such networks. The saturation throughput region corresponding to optimal $p_i$ values for $p_i$-CSMA for the special case that all nodes are in range of each other (i.e., a complete graph) was also independently characterized in~\cite{gai2011saturation}.

Ni and Srikant~\cite{ni2009distributed} as well as Jiang and Walrand~\cite{jiang2009distributed} show that, together with appropriate queue-aware policies, CSMA-based networks can achieve the maximum possible network throughput for a given set of flows. They consider a conflict graph G representation, where the set of transmitting links forms an independent set, and the full throughput region corresponds to the convex hull of these independent sets. These works analyze the transmission patterns generated by CSMA through Markovian models and show that the transmission parameters ($p_i$) of links can be adjusted to reach any point within the interior of the throughput region, if the packet lengths are sufficiently long. 

Subramanian and Leith~\cite{subramanian2011delay} provide a lower bound on throughput as well as an upper bound on queue length, where the latter is solely influenced by the local contention encountered by each link. For networks characterized by a conflict graph with a bounded degree, their work demonstrates the existence of a subset within the capacity region such that when arrival rates fall within this reduced-rate region, the mean delay can be bounded independently of the network size. 

König and Shafigh~\cite{konig2022multi} introduce an elegant Markov renewal process framework to analyze multichannel ALOHA and CSMA systems in continuous time. Their approach yields closed‑form expressions for asymptotic throughput and derives exponential upper bounds for rare events, assuming that all nodes contend simultaneously.

Zhang et al.~\cite{zhang2025distributed} propose a distributed CSMA/CA MAC protocol for RIS-assisted networks that integrates channel contention, RIS CSI acquisition, and RIS-assisted transmission via an optimal stopping framework. They derive closed-form threshold expressions that balance CSI overhead and effective transmission time, resulting in a low-complexity (O(L)) scheme that outperforms conventional strategies under various network conditions.


Singh and Kumar~\cite{singh2021adaptive} propose an adaptive CSMA protocol for decentralized scheduling in multihop wireless networks with end-to-end deadline constraints. By tuning transmission probabilities based on local state information, their approach achieves near-optimal timely-throughput without centralized control, providing a robust framework for managing deadline-sensitive traffic in interference-limited networks.

Tripathi et al.~\cite{tripathi2023fresh} propose Fresh-CSMA, a distributed protocol that minimizes AoI in single-hop wireless networks by fine-tuning backoff timers. Under ideal conditions, it mimics a centralized max-weight scheduler, and in realistic settings it reduces collisions and overhead, improving freshness and throughput.

Sammour et al.~\cite{sammour2024intelligent} introduce Intelligent CSMA/CA (ICSMA/CA), a deep reinforcement learning-based mechanism designed to enhance CSMA/CA performance in dense Wi-Fi networks. By dynamically adapting the duration of the backoff using local network state information, ICSMA/CA improves channel capacity utilization, reduces access delays, and maintains fairness between nodes, outperforming traditional CSMA/CA in high load scenarios.

Fan et al.~\cite{fan2021towards} present an analytical model to address saturation throughput disparity in ad hoc networks based on directional CSMA/CA. By introducing a four-dimensional Markov chain, they capture the effects of directional antenna-related issues such as deafness and hidden terminal problems, which significantly impact throughput fairness. Their model distinguishes between instantaneous and persistent collisions, providing accurate saturation throughput estimates validated against simulation results.

These prior works either focus on characterizing the rate region and optimized transmission probabilities (which result in pareto optimal saturation throughput values) for the special case where all nodes are in range of each other, or on showing how to optimize rates in a CSMA network with a general conflict graph. \emph{To our knowledge, exact numerical computation of saturation throughput values for $p_i$-CSMA for a given set of access probabilities for a general conflict graph has been an open problem until this work.}




\section{Problem Definition}





\begin{figure}[h!]
    \centering
    \includegraphics[width=0.5\linewidth]{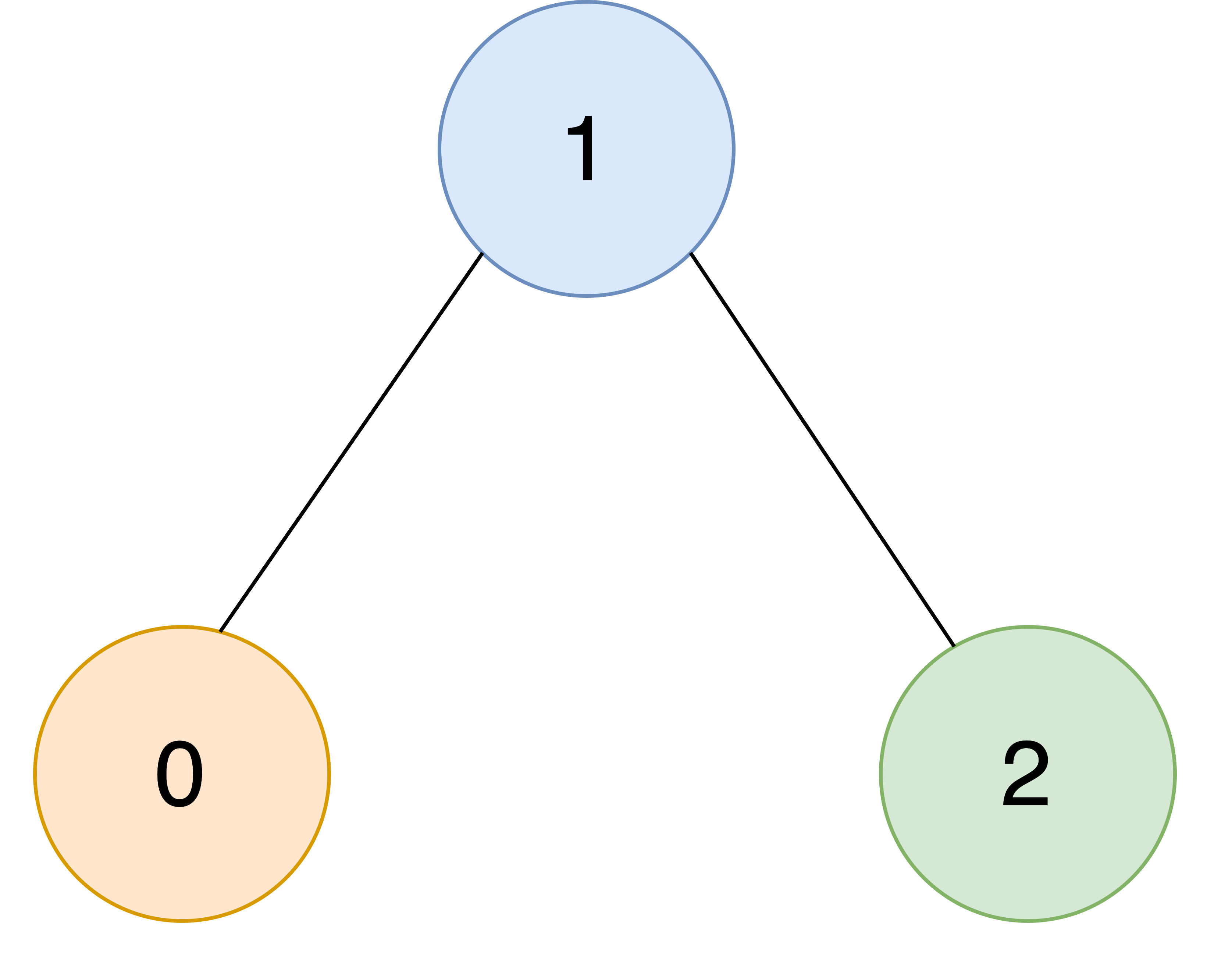}
    \captionsetup{font=small}
    \caption{3-Node Non-Complete Topology.}
    \label{fig:3nodes_non_com}
\end{figure}

\begin{figure}[h!]
    \centering
    \includegraphics[width=1\linewidth]{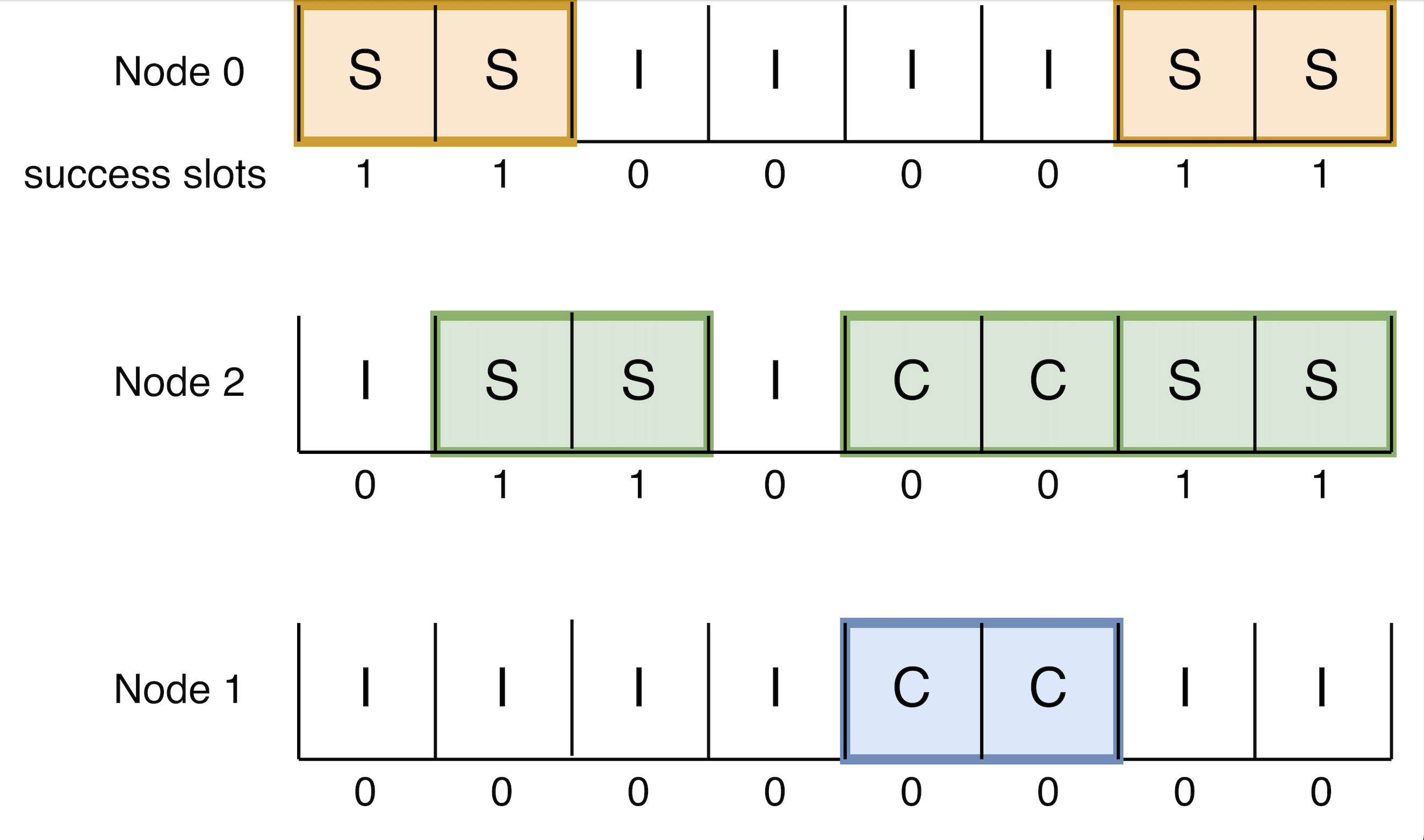}
    \captionsetup{font=small}
    \caption{Idle, Successful Transmission, and Collision outcomes in a p-persistent CSMA time slot for T=2.}
    \label{fig:slot_outcomes}
\end{figure}

We model the wireless network using a conflict graph \( G = (V, E) \), where each vertex \( v_i \in V \) represents a node in the network, and an edge \( (v_i, v_j) \in E \) indicates that nodes \( i \) and \( j \) cannot transmit simultaneously due to mutual interference. Time is divided into discrete slots of equal duration. Given that a node (which is assumed to be saturated, that is, always has a packet ready to send) attempts a transmission during a particular slot, the node will be busy for the next \(T\) slots. If a neighbor transmits during a time-slot in which a node intends to transmit, the node will defer transmission and remain idle until all neighboring nodes are silent. However, if both the node and its neighbors refrain from transmitting, it will idle for only one slot before attempting transmission again. Given that a node \(i\) is eligible to transmit in a slot, it will transmit with probability \( p_i \) independently of other nodes. 
A successful transmission for a node happens in a slot when that node transmits, and none of its neighbors transmit during the slot. 
Figure~\ref{fig:3nodes_non_com} illustrates a 3-node non-complete topology while figure \ref{fig:slot_outcomes} presents the resulting idle, successful transmission, and collision outcomes, demonstrating how node interactions and network topology influence transmission dynamics. For each node, each colored bar represents a single transmission attempt (labeled either 'S' for success or 'C' for collision), while the uncolored idle slots are labeled with 'I'. The bit streams below represent whether the transmissions are successful or not ('1' for success, '0' for collision or idle slots). The saturation throughput \( S_i \) for node \( i \) is defined as the long-term average successful transmission rate of node \( i \). The problem we are concerned with in this paper is: \emph{given the access probabilities $p_i$ for each node, compute the saturation throughput for each node.}

\section{Renewal Theory Expressions and Shortcomings}

Understanding channel access dynamics in wireless networks is crucial for evaluating the performance of medium access control protocols, such as CSMA.
One of the foundational approaches used in this context is renewal theory, which allows for an analytical characterization of the system by treating it as a renewal process~\cite{ross2014introduction}.

Let \( S_i \) denote the normalized saturation throughput for user \( i \), defined as the asymptotic time-averaged proportion of the time that the channel is used for successful packet transmissions by user \( i \).

In the special case that all nodes are in the range of each other, we can analyze the system as a renewal system by defining a renewal frame that begins whenever all the nodes are free to transmit. By applying the renewal reward theorem, the saturation throughput can be expressed as:

\begin{equation}
\begin{aligned}
S_i &= \frac{\mathbb{E}[\text{successful transmission time for } i \text{ per epoch}]}{\mathbb{E}[\text{epoch duration}]} \\
&= \frac{p_i \prod_{j \neq i} (1 - p_j) T}{\sigma \prod_{i}(1 - p_i) + \left( 1 - \prod_{i}(1 - p_i) \right) T}
\end{aligned}
\label{eqREN1}
\end{equation}

This formula works effectively in complete graphs because when any node becomes busy, the entire system is affected. 
\\
We can extend and approximate the formula for a non-complete graph using,
\small
\begin{align}
S_i = \frac{p_i \prod_{j \in N(i)} (1 - p_j)T}
{\sigma (1 - p_i) \prod_{j \in N(i)} (1 - p_j) 
+ \left(1 - \prod_{j \in N(i)} (1 - p_j)\right)T },
\label{eqREN2}
\end{align}
\normalsize
where \( N(i) \) represents the set of neighbors of node \( i \). This formulation is an approximation, constrained by the unpredictable nature of renewal events in non-complete graphs. It presumes independent transmit probabilities and oversimplifies the intricate, mutually influential interactions among nodes, neglecting complex interdependencies and temporal correlations. Consider, for example, figure~\ref{fig:3nodes_non_com}. Imagine a scenario in which nodes 0 and 2 alternate transmissions until time \( t \). At time \( t+1 \), no transmission occurs, which causes a system renewal precisely at that time. This leads to unpredictable renewal times and frame durations. The broad range of possible outcomes contributes to the highly variable frame length. Moreover, this example only illustrates the expected frame length; calculating the number of successful transmissions for each user further complicates the application of renewal theory. We shall see that these expressions are incorrect and lead to poor approximations in general networks. 

\section{Proposed Exact Markov Chain-based Approach}


In the following, for simplicity, we assume that the idle slot duration $\sigma$=1 (this does not affect generality, as it could alternatively be interpreted as measuring the transmission duration of the packet $T$ in units of $\sigma$). First, we introduce the following lemma with respect to Markov processes.
\begin{lemma}\label{lemma:markov_lemma}
    Let \( X_0, U_1, U_2, \ldots \) be independent discrete random variables. Consider the process \( \{X_n\}_{n \geq 0} \) defined by:

\[
X_{n+1} = h_n(X_n, U_{n+1}) \quad \text{for } n \geq 0,
\]

where each \( h_n \) is a deterministic function. Then \( \{X_n\}_{n \geq 0} \) is a Markov chain.
\begin{proof}
    See~\cite{norris1998markov}.
\end{proof}
\end{lemma}

We define the state of the system at time \( t \) as \( S(t) = (a_0(t), a_1(t), \dots, a_{n-1}(t)) \), where:
\[
    a_i(t) = \max\{T - \text{(Time slots since node } i \text{ last transmitted)}, 0\}
\]
In this formulation, \( a_i(t) \) represents the remaining time slots until the next possible transmission of node \( i \).

The node \( i \) can transmit during the time slot \( t \) if and only if \( a_i(t) = 0 \) and \( a_j(t) = 0 \) for all neighboring nodes \( j \) of \( i \).

\par The system dynamics are governed by the following function:
\begin{align}\label{eqn:f_def}
    S(t+1) = f(S(t), tx(t)),
\end{align}
where \(\{ tx(t): t \geq 1\} \) is a collection of i.i.d. \( n \)-dimensional binary random vectors, and \( tx_j(t) \) follows a Bernoulli distribution with parameter \( p_j \).

The function \( f \) determines the next state using the current state $S(t)$ and the random vector $tx(t)$. The procedure encompassed by $f$ can be described by first evaluating the current state \( S(t) \) to identify which nodes are eligible to transmit. For the nodes that can transmit, \( f \) examines the transmission vector \( tx(t) \) to decide their transmission status. In particular, if node $i$ is eligible to transmit during time slots $t$ and $tx_i(t) = 1$, then node $i$ transmits during time-slot $t$. If node $i$ is not eligible to transmit, then $tx_i(t)$ is ignored. Now, the function $f$ calculates $S(t+1)$. In particular,
\small
\begin{align}
    S_i(t+1) =
    \begin{cases}
        T-1, & \text{if node } i \text{ transmitted}, \\
        \max\{0, S_i(t)-1\}, & \text{otherwise}.
    \end{cases}
\end{align}
\normalsize

Considering the process \( \{S(t)\}_{t \geq 0} \), since the collection \( tx(1), tx(2), \dots \) is i.i.d. and independent of \( S(0) \), we can conclude from Lemma~\ref{lemma:markov_lemma} that \( \{S(t)\} \) forms a Markov chain.
We have defined the state space and the transition probabilities, which lead to a Markov chain \(\{S(t)\}\). The Markov chain reaches a steady state as \( t \to \infty \), where the probability distribution over states converges to the stationary distribution \(\vec{\pi}\). In the following, we provide an explicit solution to find \(\vec{\pi}\) for the special case $T = 2$.

\begin{theorem}
    For $T=2$, the stationary distribution $\vec{\pi}$ is given by
    \begin{align}\label{eqn:stat_dist}
    \pi_{(s_1, s_2, \ldots, s_n)} = \frac{\prod_{i \in \mathcal{G}_1}p_i \prod_{i \in \mathcal{G}_2} (1-p_i)}{Z},
\end{align}
where,
\begin{align}
    \mathcal{G}_1 = \{ i: s_i = 1\}
\end{align}
and
\begin{align}
    \mathcal{G}_2 = \{ i: s_i = 0, \text{ and } s_j = 1 \text{ for some neighbour of } j\}.
\end{align}
\begin{proof}    Consider two states $s^1,s^2 \in \{0,1\}^n$. We prove that the detailed balance equations given by,
\begin{align}
    \pi_{s^1}p_{s^1,s^2} = \pi_{s^2}p_{s^2,s^1} 
\end{align}
We define the sets as in \eqref{eqn:stat_dist}:
\begin{align}
    \mathcal{G}^1_1 = \{i \mid s^1_i = 1\}
\end{align}
\begin{align}
    \mathcal{G}^1_2 &= \{i \mid s^1_i = 0 \text{ and } s^1_j = 1 \text{ for a neighbor } j\} \\
    \mathcal{G}^2_1 &= \{i \mid s^2_i = 1\} \\
    \mathcal{G}^2_2 &= \{i \mid s^2_i = 0 \text{ and } s^2_j = 1 \text{ for a neighbor } j\}
\end{align}

We consider the following cases:

 \textbf{Case 1:}
\( \mathcal{G}^1_1 \cap \mathcal{G}^2_1 \neq \emptyset \).
In this case, \( p_{s^1,s^2} = p_{s^2,s^1} = 0 \), so the condition is satisfied.

\textbf{Case 2:} \( \mathcal{G}^1_1 \cap \mathcal{G}^2_1 = \emptyset \) but there exists a neighbor conflict, where either \( \mathcal{G}^1_2 \cap \mathcal{G}^2_1 \neq \emptyset \) or \( \mathcal{G}^1_1 \cap \mathcal{G}^2_2 \neq \emptyset \). In this case, there is an index \( i \) such that \( s_i \) and a neighbor \( j \) of \( i \) have conflicting states, resulting in \( p_{s^1,s^2} = p_{s^2,s^1} = 0 \).

\textbf{Case 3:} No neighbor conflicts: \( \mathcal{G}^1_1 \cap \mathcal{G}^2_1 = \emptyset \), \( \mathcal{G}^1_2 \cap \mathcal{G}^2_1 = \emptyset \), and \( \mathcal{G}^1_1 \cap \mathcal{G}^2_2 = \emptyset \). Here, we proceed with the calculations for \( \pi_{s^1} p_{s^1,s^2} \) and \( \pi_{s^2} p_{s^2,s^1} \).

Notice that now $\mathcal{G}^1_2, \mathcal{G}^2_1, \mathcal{G}^1_1$ are pairwise disjoint. 
\begin{align}
    p_{s^1,s^2} = \prod_{i \in \mathcal{G}^2_1}p_i\prod_{i \in (\mathcal{G}^1_1\cup \mathcal{G}^1_2 \cup \mathcal{G}^2_1)^c}(1-p_i)
\end{align}
and
\begin{align}
    \pi_{s^1} = \prod_{i \in \mathcal{G}^1_1}p_i\prod_{i \in \mathcal{G}^1_2}(1-p_i)
\end{align}
Hence,
\begin{align}
    \pi_{s^1}p_{s^1,s^2} =  \prod_{i \in \mathcal{G}^1_1 \cup \mathcal{G}^2_1 }p_i\prod_{i \in (\mathcal{G}^1_1 \cup \mathcal{G}^2_1)^c}(1-p_i)
\end{align}
Similarly, from symmetry, it can be easily seen that,
\begin{align}
    \pi_{s^2}p_{s^2,s^1} =  \prod_{i \in \mathcal{G}^1_1 \cup \mathcal{G}^2_1 }p_i\prod_{i \in (\mathcal{G}^1_1 \cup \mathcal{G}^2_1)^c}(1-p_i)
\end{align}

Thus, the detailed balance is satisfied. 
\end{proof}
\end{theorem}

Recall that the throughput \(S_i\) for a node \(i\) is defined as the long-term average rate at which successful transmissions occur for that node. For each state \(s\) and node \(i\), we define \(R_i(s)\) as the steady-state probability of node \(i\) having a successful transmission given the system is currently in state \(s\). Hence, the long-term proportion of successful transmissions of node \(i\) when the system is in state \(s\) is \( T \cdot R_i(s) \cdot \pi(s) \).

Considering all the states, the throughput for node \(i\) is the sum of the above over all possible states, weighted by the stationary probabilities. Hence, the throughput of node \(i\) is given by:
\begin{equation}
S_i = T\cdot \sum_{s} \pi(s) \cdot R_i(s)
\label{eqMarkov}
\end{equation}

For general $T$, we now introduce the algorithm (1), which iterates over all possible states and transmission vectors. For each state, we first identify the set \( X \) of nodes that can transmit. A node \( i \) can transmit in the current time slot if and only if \( a_i(t) = 0 \) and \( a_j(t) = 0 \) for all neighboring nodes \( j \). A transmission vector \( tx \) is valid for the state \( S(t) \) if \( tx_i(t) = 1 \) implies that \( i \in X \). Given the state \( S(t) \) and a transmission vector \( tx \), we can uniquely determine the next state \( S(t+1) \) using (\ref{eqn:f_def}). The probability of transitioning from \( S(t) \) to \( S(t+1) \) is simply the probability of the transmission vector \( tx \) occurs in state \( S(t) \). Applying this process, we compute the transition probability matrix of the Markov chain.

\begin{algorithm}[h!]
\caption{Throughput Calculation $(G, p, T)$}

\begin{algorithmic}[1]
\State\textbf{Inputs:} Network graph $G$, Probability vector  $p= \{p_1, p_2, \ldots, p_n\}$, Transmission duration $T$.
\State Initialize $\text{transMat} \gets \vec{0}^{\text{num\_states} \times \text{num\_states}}$
\State Initialize $\text{succMat} \gets \vec{0}^{\text{num\_states} \times \text{num\_nodes}}$
\State Define $States = \{ X \in \mathbb{N}^n \mid 0 \leq x_i < T, \forall i \}$
\State Define $txDecisions = \{tx \in \{0,1\}^n\}$

\For{each state $s_t \in States$}
    \State $X \gets$ set of nodes eligible to transmit in $s_t$
    \For{each $tx \in txDecisions$}
        \If{$tx$ is valid for $s_t$}
            \For{each node $i$}
                \If{node $i$ transmits}
                    \State $s_i(t+1) \gets T-1$
                \Else
                    \State $s_i(t+1) \gets \max(0, s_i(t) - 1)$
                \EndIf
            \EndFor
            \State Initialize $succP \gets 1$
            \For{each $i \in X$}
                \If{$tx[i] = 1$}
                    \State $succP \gets succP \times p_i$
                \Else
                    \State $succP \gets succP \times (1 - p_i)$
                \EndIf
            \EndFor
            \State $\text{transMat}[s_t, s_{t+1}] \gets \text{transMat}[s_t, s_{t+1}] + \rlap{$succP$}$
            \For{each node $i$}
                \If{$tx_i = 1$}
                    \State $\text{succMat}[s_t, i] \gets \text{succMat}[s_t, i] + \rlap{$succP$}$
                \EndIf
            \EndFor
        \EndIf
    \EndFor
\EndFor

\State Compute stationary distribution $\pi \gets$ solve($\text{transMat}$)

\For{each node $i$}
    \For{each state $s \in States$}
        \State $\text{throughput}[i] \gets \text{throughput}[i] + \text{succMat}[s, i] \cdot \rlap{$\pi[s]$}$
        
    \EndFor
\EndFor

\State \textbf{Output:} Achieved throughput $\text{throughput}[i]$ for each node $i$.

\end{algorithmic}
\end{algorithm}

\normalsize

\begin{figure}[h!]
    \centering
    \includegraphics[width=1\linewidth]{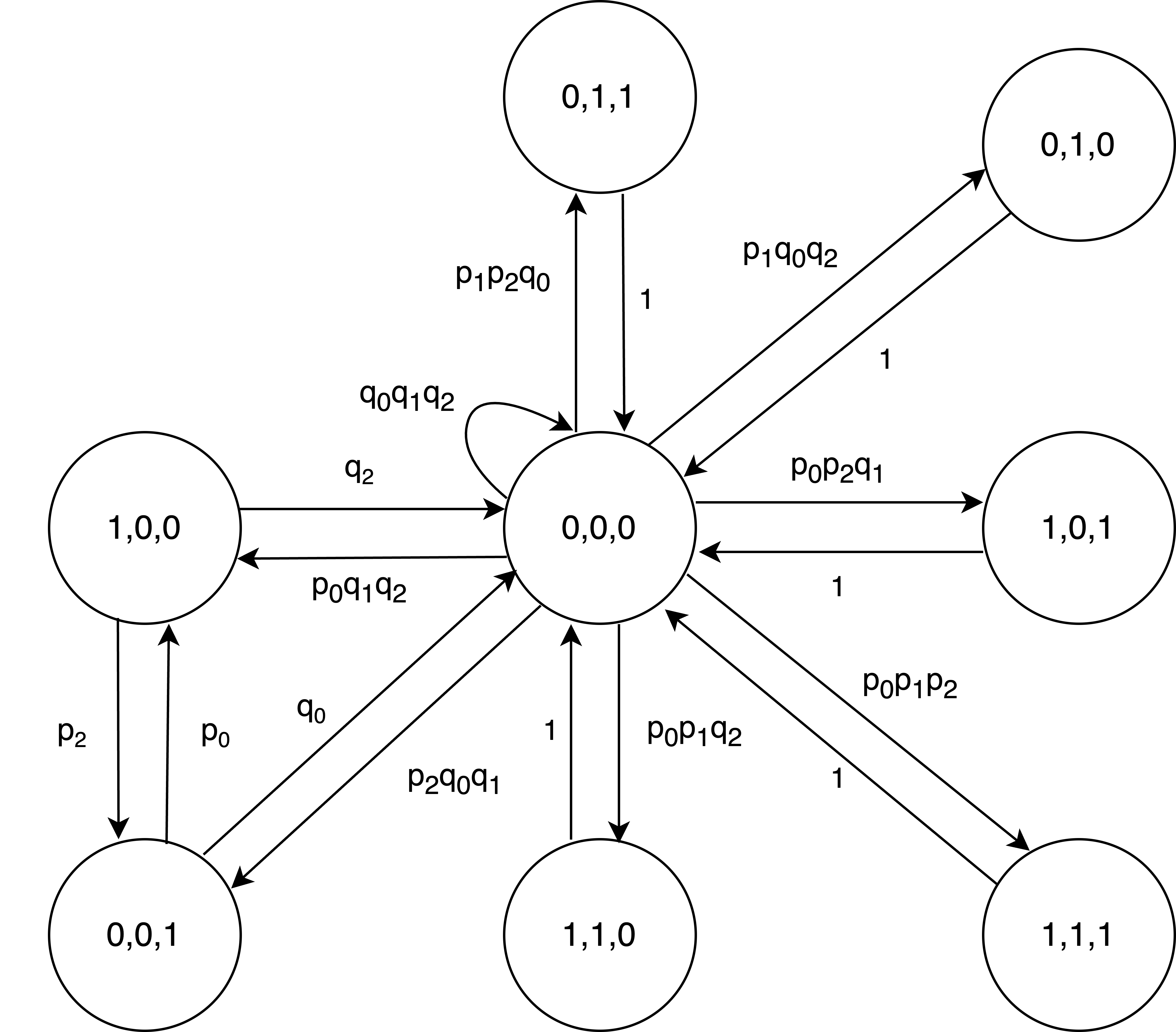}
    \captionsetup{font=small}
    \caption{State Transition Diagram of the Markov Chain for the 3-Node Non-Complete Graph with \( T = 2 \) and \( \sigma = 1 \)}
    \label{fig:MarkoveExample}
\end{figure}

As an example, consider the network topology shown in figure.\ref{fig:3nodes_non_com};  with parameters \( T = 2 \) and \( \sigma = 1 \), the corresponding Markov chain state space is defined as follows, and the Markov chain for this system is illustrated in figure~\ref{fig:MarkoveExample}.

Each state in the Markov chain is represented by the vector \( S(t) = (a_1(t), a_2(t), a_3(t)) \), where \( a_i(t) \) denotes the backoff counter for Node~\( i \) at time \( t \). Since \( T = 2 \), the possible values for \( a_i(t) \) are 0 and 1, representing the number of remaining time slots before a node can attempt to transmit again.

The following transition matrix represents the state transitions based on the nodes' transmission probabilities, where \( q_i = 1 - p_i \):

\scriptsize
\(
\begin{array}{c@{}|@{}c@{}c@{}c@{}c@{}c@{}c@{}c@{}c}
& (0,0,0) & (0,0,1) & (0,1,0) & (0,1,1) & (1,0,0) & (1,0,1) & (1,1,0) & (1,1,1) \\
\hline
(0,0,0) & q_0 q_1 q_2 & q_0 q_1 p_2 & q_0 p_1 q_2 & q_0 p_1 p_2 & p_0 q_1 q_2 & p_0 q_1 p_2 & p_0 p_1 q_2 & p_0 p_1 p_2 \\
(0,0,1) & q_0 & 0 & 0 & 0 & p_0 & 0 & 0 & 0 \\
(0,1,0) & 1 & 0 & 0 & 0 & 0 & 0 & 0 & 0 \\
(0,1,1) & 1 & 0 & 0 & 0 & 0 & 0 & 0 & 0 \\
(1,0,0) & q_2 & p_2 & 0 & 0 & 0 & 0 & 0 & 0 \\
(1,0,1) & 1 & 0 & 0 & 0 & 0 & 0 & 0 & 0 \\
(1,1,0) & 1 & 0 & 0 & 0 & 0 & 0 & 0 & 0 \\
(1,1,1) & 1 & 0 & 0 & 0 & 0 & 0 & 0 & 0 \\
\end{array}
\)
\normalsize
\newline
\newline
The stationary distribution \( \vec{\pi} \) is:
\\
\(
\vec{\pi} = [\pi_{000},\ \pi_{001},\ \pi_{010},\ \pi_{011},\ \pi_{100},\ \pi_{101},\ \pi_{110},\ \pi_{111}]
\)
\\

The long-term stationary distribution of the Markov chain gives the probability of successful transmission for each node, directly determining its throughput. Thus, the throughput for each node can be calculated as:

\begin{align*}
S_0 &= T\cdot\left(\pi_{000}(p_0 q_1 q_2 + p_0 q_1 p_2) + \pi_{001} p_0\right), \\
S_1 &= T\cdot\left(\pi_{000}(q_0 p_1 q_2)\right), \\
S_2 &= T\cdot\left(\pi_{000}(p_0 q_1 p_2 + q_0 q_1 p_2) + \pi_{100} p_2\right).\end{align*}

\[
\begin{aligned}
\pi_{000} &= \frac{1}{1 + q_1 p_2 + q_1 p_0 +p_1 + q_1 p_0 p_2}, \\[2mm]
\pi_{001} &= \frac{q_1 \, p_2}{1 + q_1 p_2 + q_1 p_0 +  p_1 + q_1 p_0 p_2}, \\[2mm]
\pi_{100} &= \frac{p_0 \, q_1}{1 + q_1 p_2 + q_1 p_0 + p_1 + q_1 p_0 p_2}, \\[2mm].
\end{aligned}
\]

Incorporating the stationary distributions of interest, as defined in equation~(\ref{eqn:stat_dist}), 
we get that the final expressions for \( S_0 \), \( S_1 \), and \( S_2 \) are :

\begin{align}
S_0 &= 2\cdot\left(\frac{p_0 q_1 q_2 + p_0 q_1 p_2 + q_1 p_0 p_2}{1 + q_1 p_2 + q_1 p_0 + p_1 + q_1 p_0 p_2}\right), \nonumber \\[2mm]
S_1 &= 2\cdot\left(\frac{q_0 p_1 q_2}{1 + q_1 p_2 + q_1 p_0 + p_1 + q_1 p_0 p_2}\right), \nonumber \\[2mm]
S_2 &= 2\cdot\left(\frac{p_0 q_1 p_2 + q_0 q_1 p_2 + p_0 q_1 p_2}{1 + q_1 p_2 + q_1 p_0 + p_1 + q_1 p_0 p_2}\right).
\label{eqn:3rates}
\end{align}
\newline

\section{Application to Network Utility Maximization}

In this section, we focus on optimizing the transmit probabilities to maximize a weighted sum of a utility function $U()$ of the saturation throughput values \( S_i \). In particular, given non-negative weights \( \alpha_1, \alpha_2, \dots, \alpha_n \), 
we aim to find the optimal probabilities \( p_1, p_2, \dots, p_n \)  to solve the problem

\begin{align}
   & \max_{\{p_1, \dots, p_n\}} J(p_1, \dots, p_n) \\
   & \text{ subject to } p_i \in [0,1] \ \forall i \in \{1,2,\dots,n\}.
\end{align}
where the function $J$ is given by,
\begin{align}
     J(p_1, \dots, p_n) = \sum_{i=1}^{n} \alpha_i U(S_i(p_1, p_2, \dots, p_n))
\end{align}

Solving this problem for a given utility function $U()$ with different values of $(\alpha_1,\alpha_2,\dots,\alpha_n)$ will generate different $(S_1,S_2,\dots,S_n)$ that optimize the weighted sum utilities. Since each \( S_i \) is a function of the probabilities \( p_i \), we use gradient ascent with projections onto the feasible region to maximize \( J \) with respect to \( p_i \) while satisfying the constraints.

For the network depicted in figure~\ref{fig:3nodes_non_com}, we compute the gradient of $J$ as:
\begin{equation}
\frac{\partial J}{\partial p_i} = \alpha_1 \frac{\partial U(S_0)}{\partial p_i} + \alpha_2 \frac{\partial U(S_1)}{\partial p_i} + \alpha_3\frac{\partial U(S_2)}{\partial p_i},
\end{equation}
where $S_0,S_1,S_2$ are given in \eqref{eqn:3rates}.
We utilize gradient ascent with projections onto the feasible region to find the optimal transmit probabilities. The update rule for \( p_i \) is:

\begin{equation}
p_i^{(k+1)} = \left[p_i^{(k)} + \eta_k \frac{\partial J}{\partial p_i} \Bigg|_{p_i = p_i^{(k)}}\right]_{[0,1]},
\end{equation}

where $[x]_{[0,1]}$ denoted the projection of $x$ onto $[0,1]$ and \( \eta_k > 0 \) is the learning rate. In particular,
\begin{align}
    [x]_{[0,1]} = \min\left( \max\left( x, 0 \right), 1 \right)
\end{align}
The gradient ascent algorithm will converge to a local optimum for the above problem.

\begin{figure*}[ht!]
    \centering
    \includegraphics[width=17cm, height=5.5cm]{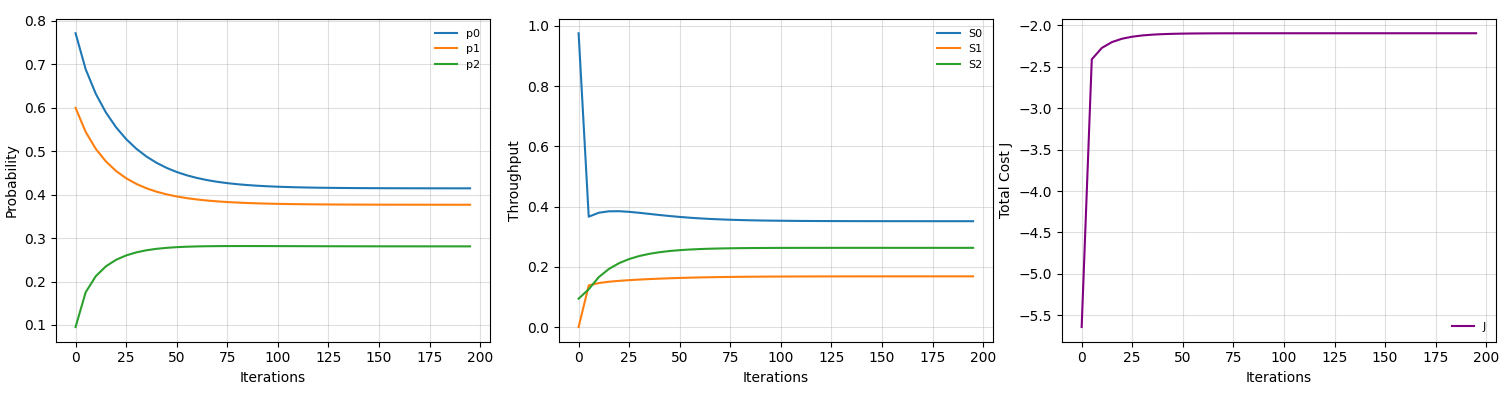}  
    \captionsetup{font=small}
    \caption{Illustrating convergence of Utility optimization for the topology in figure \ref{fig:3nodes_non_com}}
    \label{fig:convergence}
\end{figure*}

\section{Numerical Results}
\begin{figure}[htb]
    \centering
    \includegraphics[width=1\linewidth]{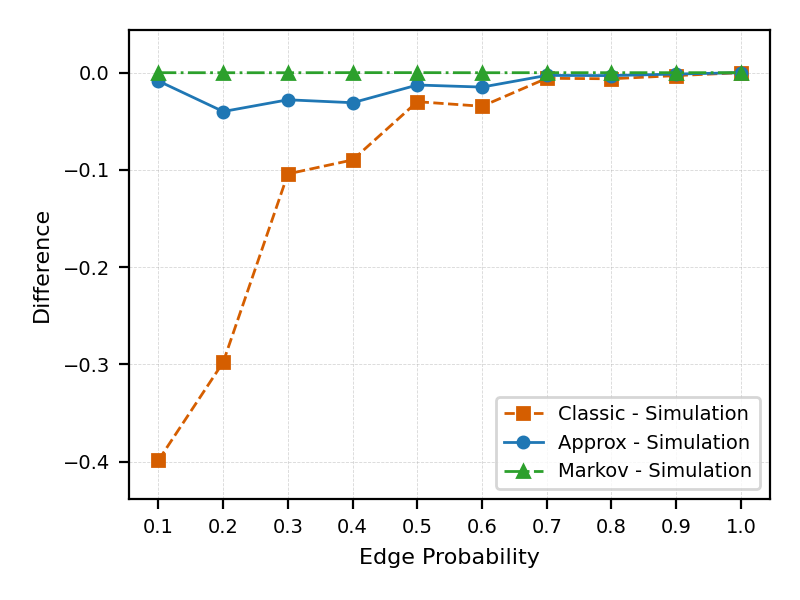}
    \captionsetup{font=small}
    \caption{ Theoretical Models and simulation differences across edge probabilities}
    \label{fig:differences}
\end{figure}

In this section, we conducted a comparative analysis of two distinct analytical approaches—Renewal Theory described in equations (\ref{eqREN1}) and (\ref{eqREN2}) and a Markov Chain Model in equation (\ref{eqMarkov}) against simulation data.
We employed the Erdős-Rényi \cite{bhamidi2014scaling} model to generate random graphs with 10 nodes, representing levels of connectivity from sparse to fully connected. We can see from table~\ref{tab:comparison} and figure~\ref{fig:differences} that our Markov chain-based calculation matches the simulated throughput accurately while the classic renewal theory equation (\ref{eqREN1}) and extended renewal theory approximation equation (\ref{eqREN2}) perform poorly especially in sparser graphs. 

\begin{table}[h!]
\centering
\renewcommand{\arraystretch}{1.1} 
\small 
\begin{tabular}{p{1.3cm} p{1.3cm} p{1.3cm} p{1.3cm} p{1.3cm}}
\hline
\textbf{Edge Prob.} & \textbf{Sim. Thpt} & \textbf{Renewal Classic} & \textbf{Renewal Approx} & \textbf{Markov} \\
\hline
0.1 & 0.39881 & 0.00030 & 0.39022 & 0.39880 \\
0.2 & 0.29799 & 0.00025 & 0.25805 & 0.29795 \\
0.3 & 0.10607 & 0.00198 & 0.07820 & 0.10601 \\
0.4 & 0.09003 & 0.00032 & 0.05914 & 0.09008 \\
0.5 & 0.03117 & 0.00133 & 0.01849 & 0.03121 \\
0.6 & 0.03517 & 0.00063 & 0.02036 & 0.03512 \\
0.7 & 0.00638 & 0.00060 & 0.00357 & 0.00638 \\
0.8 & 0.00795 & 0.00166 & 0.00496 & 0.00795 \\
0.9 & 0.00342 & 0.00042 & 0.00201 & 0.00342 \\
1.0 & 0.00059 & 0.00059 & 0.00059 & 0.00059 \\
\hline
\end{tabular}
\captionsetup{font=small} 
\caption{Throughput Comparison by Edge Probability: Simulation and Theoretical Models}
\label{tab:comparison}
\end{table}

\begin{figure}[htb]
    \centering
    \includegraphics[width=1\linewidth]{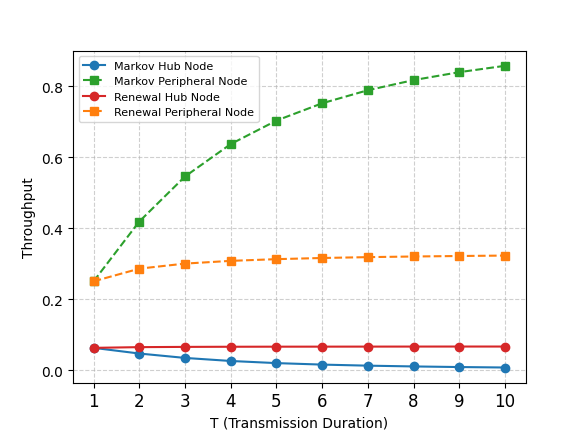}
        \captionsetup{font=small}
    \caption{Impact of Transmission Duration T on throughput in a star topology: Markov vs. Renewal approximations (Renewal underestimates by up to 60\%)}
    \label{fig:T}
\end{figure}

\begin{figure}[h!]
    \centering
    \includegraphics[width=1 \linewidth]{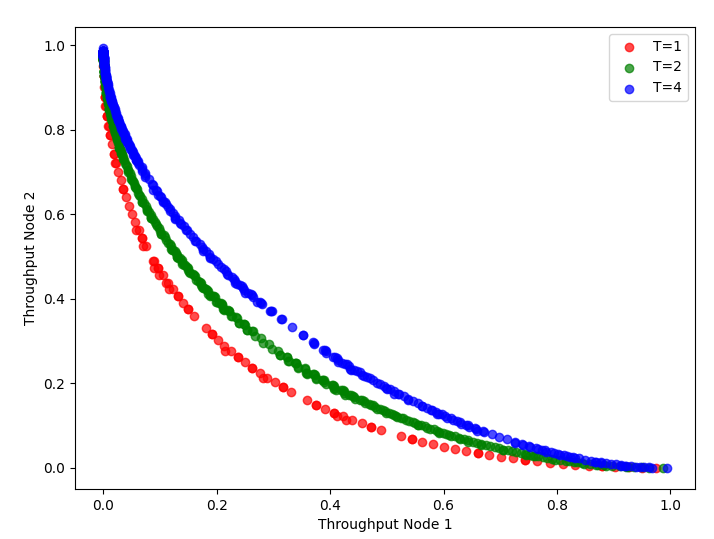}
    \captionsetup{font=small}
    \caption{The boundary of the saturation throughput region, illustrated for node 1 and 2 (while fixing throughput of node 3 at 0) of the topology in figure \ref{fig:3nodes_non_com}}
    \label{fig:throughputBoundary}
\end{figure}

Figure \ref{fig:T} illustrates how varying the transmission duration (T) affects throughput while keeping the network topology and transmission probability of the nodes constant. As T increases, the throughput of the hub node declines due to higher contention, whereas the peripheral nodes benefit from reduced collisions. The Markov model accurately captures these dynamics, showing a significant increase in throughput for peripheral nodes, whereas the Renewal theory approximation fails to account for these effects, underestimating throughput by 48–62\% at larger T. This highlights the importance of using the Markov model for precise CSMA network analysis and optimization.

In figure \ref{fig:throughputBoundary}, we present the saturation throughput region for two of the nodes, for the topology in figure 1. As the value of T increases (indicating longer packet transmissions), the throughput region
expands due to more efficient use of the communication
channel.

Finally, in figure~\ref{fig:convergence}  we illustrate network utility maximization using the saturation throughput computation introduced in this paper. For the 3-node topology of figure~\ref{fig:3nodes_non_com} (for which the throughput equations are given in equation~\ref{eqn:3rates}). assuming a logarithmic utility function (i.e. $U() = log()$), for weights $\alpha = [0.6 , 0.6, 0.3]$, it shows how the probabilities, saturation throughput values and the total utility being maximized converge over iterations of the gradient ascent.

\section{Conclusions}

In this paper, we developed a rigorous framework to compute the exact saturation throughput of heterogeneous $p$-persistent CSMA in wireless networks with arbitrary conflict graph topologies, moving well beyond the conventional assumption of fully connected networks. By formulating the problem within a global Markov chain framework, we derived closed‐form expressions for the steady‐state probabilities governing successful transmissions, which accurately capture a wide range of realistic scenarios with varying contention levels. Our novel computational approach not only shows that commonly used throughput approximations can be significantly inaccurate for general topologies but also provides exact results for any network configuration and set of transmit probabilities. In the special case where the packet length is twice the slot length, we obtained a more compact expression that serves as a consistency check for the general framework while offering deeper insight into practical performance trends. Furthermore, we demonstrated how to optimize transmission probabilities to maximize weighted sums of utility functions, thereby illustrating the potential of our method for enhancing overall network performance.

Looking ahead, we are interested in exploring the extension of this work in several promising directions. We would like to investigate how to generate more computationally tractable approximations for large‐scale networks. Adaptive or machine learning‐based algorithms could be employed to adjust transmit probabilities in real time, complementing our current methods with more flexible and responsive control mechanisms. We would also like to explore how to incorporate dynamic traffic models with finite arrival rates and queueing (going beyond the saturation / backlogged-users setting), as well as factors such as capture effects and fading, to further enhance the realism of our framework. Advanced PHY-layer features, such as multi-user MIMO, can also be integrated to capture more intricate interference relationships. Finally, the development of distributed network control schemes that take advantage of local information could approximate global optimality in a scalable manner.

\bibliographystyle{IEEEtran}
\bibliography{new_main.bib}

\end{document}